\documentclass[sigconf,nonacm=true]{acmart}

\usepackage{pdfpages}
\usepackage{amsfonts}
\usepackage{pifont}
\usepackage{colortbl}
\usepackage{xcolor}
\usepackage{microtype}
\usepackage{multirow, booktabs}
\usepackage{adjustbox}
\usepackage{enumitem}
\usepackage{float}
\usepackage{bm}
\usepackage{bbm} % \mathbbm
\usepackage{amsmath}
\usepackage{graphicx}
\usepackage{hyperref}
\usepackage{soul}
\usepackage[labelformat=simple]{subcaption}
\usepackage{color}
\usepackage{fancyhdr}

\def\model/{PBNR}

\newcommand{\etal}{\textit{et al.}}
\begin{document}
\title{PBNR: Prompt-based News Recommender System}

\author{Xinyi Li}
\affiliation{%
\institution{Northwestern University, IL, US}
\city{}
\state{}
\country{}
}
\email{xinyili2024@u.northwestern.edu}

\author{Yongfeng Zhang}
\affiliation{%
\institution{Rutgers University, NJ, US}
\city{}
\state{}
\country{}
}
\email{yongfeng.zhang@rutgers.edu}

\author{Edward C. Malthouse}
\affiliation{%
\institution{Northwestern University, IL, US}
\city{}
\state{}
\country{}
}
\email{ecm@northwestern.edu}
%% article.
\begin{abstract}

Online news platforms often use personalized news recommendation methods to help users discover articles that align with their interests. These methods typically predict a matching score between a user and a candidate article to reflect the user's preference for the article. Some previous works have used language model techniques, such as the attention mechanism, to capture users' interests based on their past behaviors, and to understand the content of articles. However, these existing model architectures require adjustments if additional information is taken into account. Pre-trained large language models, which can better capture word relationships and comprehend contexts, have seen a significant development in recent years, and these pre-trained models have the advantages of transfer learning and reducing the training time for downstream tasks. Meanwhile, prompt learning is a newly developed technique that leverages pre-trained language models by building task-specific guidance for output generations. To leverage textual information in news articles, this paper introduces the pre-trained large language model and prompt-learning to the community of news recommendation. The proposed model \emph{prompt-based news recommendation} (\model/) treats the personalized news recommendation as a text-to-text language task and designs personalized prompts to adapt to the pre-trained language model --- text-to-text transfer transformer (T5). Experimental studies using the Microsoft News dataset show that \model/ is capable of making accurate recommendations by taking into account various lengths of past behaviors of different users. \model/ can also easily adapt to new information without changing the model architecture and the training objective. Additionally, \model/ can make recommendations based on users' specific requirements, allowing human-computer interaction in the news recommendation field, which is not possible with existing news recommendation models.

\end{abstract}

\maketitle
%-------------------------------------------

\section{Introduction}

The newspaper industry has experienced a steady and steep decline over the past decade in part because traditional, ad-supported revenue models are no longer viable. There have been widespread layoffs and closures, resulting in ‘ghost newspapers’ and ‘news deserts’, where almost 200 out of 3,143 counties in the U.S have been left with no daily newspaper and 1,540 counties with only one weekly newspaper \cite{abernathy2018expanding}. The demise of local newspapers is not only a commercial problem, but also a public and social problem. Communities without news organizations have seen an increase in government spending due to a lack of accountability \cite{gao2019municipal}. Citizens who consume less news are unable to evaluate elected officials and are less likely to participate in voting. Reading news is one way for people to gain knowledge and to become more open-minded. Online platforms such as Google News and Microsoft News are attracting users to read news online \cite{wu2020mind}. However, in the current information-overloaded society, it is difficult for users to find news articles of interest from the massive set of news articles published each day \cite{lian2018towards}. Therefore, it is important to design news recommendation systems (RS) to find articles of interest for users.

A news recommender system typically involves three fundamental tasks: analyzing users' interests based on their past behaviors, comprehending news content by considering its contextual information, and predicting a user's matching score with candidate articles for personalized ranking \cite{wu2023personalized}. Beyond these tasks, news RS should also create diverse sets of story recommendations \cite{abdollahpouri2021toward}. News articles contain rich textual information, including their titles, bodies, and topics, making language model techniques like Gated Recurrent Unit (GRU) \cite{cho2014learning}, Long-short Term Memory (LSTM) \cite{graves2012long}, Convolutional Neural Network (CNN) \cite{chen2015convolutional}, and attention mechanisms \cite{vaswani2017attention} popular choices for modeling users' interests and comprehending article content \cite{an2019neural, wu2022news, wu2019neural}. 

There have recently been significant developments in pre-trained language models that can be used across various language tasks. These models can transfer knowledge from one task to another without extensive additional training, making them useful for fine-tuning for specific domains with little data compared to training a model from scratch. GPT-3 \cite{brown2020language}, BERT \cite{devlin2018bert}, and RoBERTa \cite{liu2019roberta}, are popular pre-trained language models that demonstrated impressive performance in natural language processing tasks. However, these models are designed for language-related tasks, which poses a challenge in utilizing them directly for recommendation tasks. Most existing news RS, however, still rely on techniques employed by language models, such as Gated Recurrent Unit (GRU) \cite{cho2014learning}, Convolutional Neural Network (CNN) \cite{chen2015convolutional}, and the attention mechanism \cite{vaswani2017attention}. Furthermore, these models are usually large and complex, which makes it impossible to modify their structures. To address this limitation and leverage the pre-trained language models, prompt learning \cite{jin2021good}, which provides specific prompts to guide the output generation, has been introduced. Prompt learning has proven to be an effective approach for various tasks, such as text classification and question answering, and additionally, prompt learning makes it possible to generate responses by considering the interactions with users. 

Motivated by the power of pre-trained language model and prompt learning, this paper presents a novel news recommendation model, named \model/ (Prompt-based News Recommendation), that treats the personalized news recommendation task as a text-to-text language task. The personalization comes from the description of users' past reading behaviors and articles in the designed prompts. In summary, the key contributions in this work are: 
\begin{itemize}
\item We introduce \model/, a novel approach that predicts a user's preference for an article by applying personalized prompts that model the user's past behavior and article information. Unlike the existing deep neural news recommendation methods, \model/ allows various history lengths for different users throughout the training process. To the best of our knowledge, in the community of news RS, we are the first to employ prompt learning and directly treat the news recommendation as a text-to-text language task.
\item We propose incorporating language generation loss and ranking loss during model training to enhance the language model's performance on the recommendation task. 
\item We demonstrate \model/'s flexibility in integrating additional text information to improve recommendation performance or trigger different recommendation tasks. The model is not only evaluated on the ranking performance but also on the diversity of recommended topics and the trade-off between them.
\item We investigate the potential for controlling news recommendations based on individual user requirements, which can enhance user experiences, improve human-computer interaction, and to some extent improve the interpretability of news RS with the help of personalized prompt learning. This is also the main advantage that distinguishes \model/ from existing news RS.
\end{itemize}

This work is organized as following: we briefly summarize relevant literature in section \ref{section:related_work}, describe our methodology in section \ref{section:method}, and present our experimental findings in section \ref{section:experiment}. Lastly, we make a conclusion and discuss future research directions in section \ref{section:conclusion}.

\section{Related Works}\label{section:related_work}
\textbf{Sequential News Recommendation.} Sequential news recommendation methods predict a user's preference for a candidate article based on the user's previous reading behavior. Since news articles are items with rich textual information, language techniques are often utilized to extract useful information from news contexts and understand users' interests \cite{an2019neural, wu2022news, wu2019neural}. Okura \etal\ \cite{okura2017embedding} propose using a denoising autoencoder to study news representations and use a GRU network to model users' interests. An \etal\ \cite{an2019neural} adopt CNN and the attention mechanism to learn a news representation from its title, topic and subtopic; learn a user's short-term representation using a GRU network; and learn a user's long-term representation using his/her ID embedding. The NRMS model proposed by Wu \etal\ \cite{wu2019neural2} studies a news representation from its title using a word-level, multi-head, self-attention and additive word-attention network, and studies a user's interest using a multi-head, self-attention network with the given historical clicked news sequence by a user. Wu \etal\ \cite{wu2019neural3} also propose a neural news RS approach that studies news representation using an attentive multi-view network. Besides adopting various language models to better represent users and articles, An \etal \ \cite{an2019neural} suggest not to only focus on a user's short-term interest from his/her past behavior but also study the user's long-term interest from his/her ID embedding, and Wu \etal\ \cite{wu2022news} suggest being aware of temporal diversity when modeling the match between a user and an article. All of these deep news RS highly rely on the thriving of language techniques, but they are typically trained from scratch. In contrast, \model/ treats news RS as a text-to-text language task directly, and instead of building an encoder-decoder architecture on our own, \model/ utilizes prompt learning to adapt to the pre-trained language model T5 \cite{raffel2020exploring}. Moreover, if further information is available, existing news RS models may require architectural modifications, whereas \model/ can simply integrate such information into its prompts without any need to modify its model architecture.

\textbf{Pre-trained Language Models and RS.} The development of language models can advance news RS. Pre-trained language models such as BERT \cite{devlin2018bert} and GPT \cite{radford2018improving} have been proposed and are trained on large-scale datasets, allowing them to learn general textual knowledge, and they can be easily and accurately adapted to other downstream tasks. However, modifying the structure or retraining a pre-trained model is not practical. Therefore, prompt learning \cite{jin2021good}, which designs task-specific prompts to guide the model outputs, has been introduced. Different from fine-tuning a pre-trained model on a downstream task without any prompt, the customized task-specific prompts for prompt learning leverage the knowledge learned by pre-trained language models, and can even be adapted to other domains.

Motivated by the effectiveness of pre-trained language models and prompt learning techniques, RS researchers tend to formulate recommendation as a language task. Zhang \etal\ \cite{zhang2021language} convert the item-based recommendation to a text-based cloze task by modeling a user's historical interactions to a text inquiry. Li \etal\ \cite{li2022personalized} designed personalized prompt learning for explainable recommendation by treating user and item IDs as prompts. To resolve the issue that IDs are not in the same semantic space as pre-trained language models, they proposed both discrete and continuous prompt learning, training strategies like sequential tuning and recommendation as regularization. Cui \etal\ \cite{cui2022m6} propose M6-Rec, which converts a user's behavior to a text inquiry using general textual descriptions. Inspired by the T5 model \cite{raffel2020exploring} that studies a unified text-to-text generation model, Geng \etal\ \cite{geng2022recommendation} design a flexible and unified text-to-text paradigm called `Pretrain, Personalized Prompt, and Predict Paradigm' (P5) for RS. Similar to P5 \cite{geng2022recommendation}, our \model/ is also an encoder-decoder transformer that uses T5 as a backbone. However, different from P5, which relies on user IDs and item IDs \cite{geng2022recommendation} and may encounter challenges due to discrepancies between the semantic space of these IDs and that of the pre-trained language models, our approach \model/ describes users' behaviors and news textually in the designed prompts. Furthermore, to enhance the language model's performance on the recommendation task, \model/ incorporates the ranking loss and the language generation loss throughout the training. Zhang \etal\ \cite{zhang2023prompt} employ prompt learning for news recommendation but formulate it as a cloze task for the [MASK] prediction, and it is still a traditional ranking-based recommendation. In contrast, our model, \model/, is a generative recommendation model.

\section{Methodology}\label{section:method}
Our goal is to estimate a user $u$'s preference $\hat{r}_{ui}$ to a candidate article $i$, given the user's past reading behavior. In this section, we first provide a concise description of our methodology, and then describe the objective function employed by \model/ to train the model parameters. Lastly, we introduce the prompts we have developed for prompt learning for the news recommendation task. 

\subsection{Model Architecture} \label{subsection:model_architecture}
\model/ applies transformer \cite{vaswani2017attention} blocks to build an encoder-decoder framework to map the input sequence $X = \{x_1, x_2, \dots, x_n\}$, where each token $x_i$ is either a word or a subword, into an output sequence $Y = \{y_1, y_2, \dots, y_m\}$, and both the encoder and the decoder consist of a stack of $H$ identical layers. 

At the $\ell$-th layer of the encoder, the input tokens $X^{\ell - 1}$ from the previous layer are first transformed using a multi-head self-attention mechanism, which generates a set of attention weights for each token based on its interactions with all other tokens in the sequence. This produces 

$$MH(X^{\ell - 1})= [head_1, \dots, head_h]W^O,$$$$ head_i = Attention(X^{\ell - 1}W_i^Q, X^{\ell - 1}W_i^K, X^{\ell - 1}W_i^V).$$ 
The attention applies the scaled dot product attention $$ Attention(Q, K, V) = softmax\left(\frac{QK^{T}}{\sqrt{d}}\right)V, $$ where $Q$, $V$ and $K$ represent query, value and key of dimension $d$ respectively. The output then undergoes a residual connection and layer normalization, resulting in $O^{\ell}$, which is further passed through a position-wise feed-forward neural network to get 
$$ReLU(O^{\ell}W_{1} + b_{1})W_{2} + b_{2}.$$
The output of this feed-forward network is added to the original input tokens using a residual connection, and the resulting sequence is normalized using layer normalization to get $X^{\ell}$. In the aforementioned formulas, $W^O, W_i^Q, W_i^K, W_i^V, W_1, W_2, b_1$ and $b_2$ are model parameters.

 The decoder uses a similar architecture to generate output tokens one at a time, conditioned on the input sequence and previously generated output tokens. The decoder employs a linear transformation and a softmax layer to obtain a probability distribution over all tokens.

\begin{figure*}[t!]
\centering
\includegraphics[width=15cm]{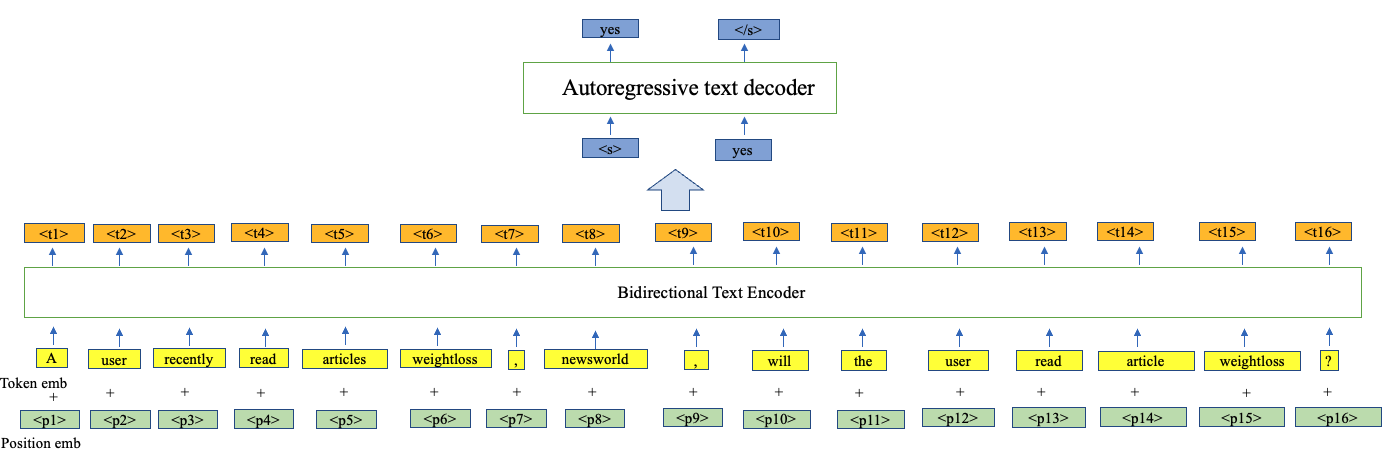}
\caption{\model/ utilizes an encoder-decoder framework, where a user's historical behavior is converted into a text inquiry and each news article is described textually, and then \model/ generates the answer to indicate a user's preference to a candidate article through an auto-regressive decoder.}
\label{fig:encoder-decoder}
\end{figure*}

\subsection{Objective Function in Modeling Training}

Each user's sequential behavior is converted into a textual input sequence, and each article is described using its textual information such as its topic, and title. Before feeding into the encoder, each input token is represented as the sum of its raw token embedding and an additional position embedding to capture the token's positional information. Given the output sequence $Y = \{y_1, y_2, \dots, y_m\}$, \model/ utilizes the auto-regressive generation, which relies on previous tokens $y_{<t}$ to estimate the probability of the next token $P_{\theta}(y_t|y<t, X)$, where $\theta$ represents all parameters in the model. The language generation loss function applied to estimate the model parameters $\theta$ for auto-regressive model is the negative log-likelihood (NLL)
$$L_{NLL} = - \sum_{t} \log P_{\theta}(y_t|y_{<t}, X).$$

The negative log-likelihood $L_{NLL}$ measures how well the language model can generate the observed output sequence; however, RS often care about how well a model ranks items for a given user, so pair-wise or list-wise training are often applied to maximize the margin between a user's preference for a positive sample $\hat{r}_{u,pos}$ and that for a negative sample $\hat{r}_{u,neg}$. To improve the language model's performance in news recommendation task, we incorporate the NLL and Bayesian Personalized Ranking (BPR) loss \cite{qi2021pp} 
$$L = (1 - \lambda)L_{NLL} + \lambda L_{BPR},$$ where $\lambda$ is a positive hyper-parameter, and BPR loss is 
$$L_{BPR} = -\sum_{(u, pos, neg)} \log(\sigma((\hat{r}_{u,pos} - \hat{r}_{u,neg}))),$$
where $\sigma(\cdot)$ denotes the sigmoid function. Since \model/ follows an encoder-decoder framework to address the recommendation task as a language problem, the estimated user's preference $\hat{r}_{ui}$ for item $i$ is represented as the probability that the generated output sequence indicates a positive sentiment. All empirical studies in this paper share the same model architecture from section \ref{subsection:model_architecture} and the integrated loss function $L$. 

\subsection{Personalized Prompts for Various News RS Tasks}
\begin{figure*}[t!]
\centering
\includegraphics[width=15cm]{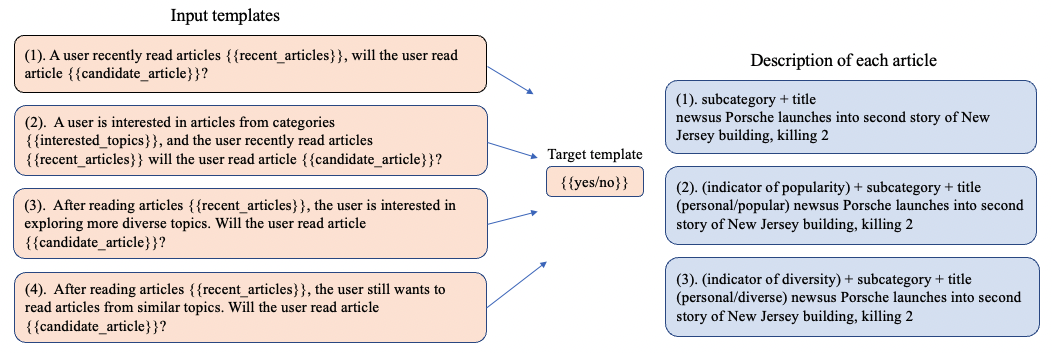}
\caption{The personalized prompts are created by designing input-target templates, wherein the relevant fields in the prompts are replaced with corresponding information from the raw data. The description of articles may differ based on specific recommendation tasks. In this study, the model denoted by \model/ ($i$-$j$) employs input template ($i$) and article description ($j$).
}
\label{fig:template}
\end{figure*}

The emergence of prompt learning make it possible to leverage the pre-trained language models for tasks in different domains. In this subsection, we discuss the prompts we have created, which consist of an input template and a target template to fit the encoder-decoder framework, as shown in Figure \ref{fig:template}. Our key focus is to treat news RS as a text-to-text language task for recommendation rather than manually constructing or exploring how to design prompts; therefore, the prompts designed in Figure \ref{fig:template} look similar to each other. 

The personalization of input templates is from the personalized depiction of a user's past behaviors and the detailed description of each article. Unlike previous studies such as Li \etal \ \cite{li2022personalized} and Geng \etal\ \cite{geng2022recommendation}, we do not use user or item IDs in our prompts. This is because IDs are not parts of the word-level description that the pre-trained language model was trained on. Furthermore, news articles have short life-cycles \cite{zhou2020s3}, implying that a large proportion of IDs become outdated over time. Therefore, relying on IDs for studying article representations might not be effective, and instead, emphasis should be placed on employing articles' contextual information.

The selection of input templates is dependent on the specific recommendation tasks at hand. For the purpose of creating personalized recommendations, input template (1) is utilized, while input template (2) is applied to include users' static attributes, such as their interests in particular topics. Moreover, input templates (3) and (4) are employed to assess the controllability of \model/, based on a user's specific requests, such as exploring more topics or reading articles from similar topics next. To ensure clarity and uniformity, a standardized target template of `{yes/no}' is adopted across all input templates, so a user's preference for an article $\hat{r}_{ui}$ used for personalized ranking is estimated as the probability that the output from the auto-regressive decoder is `yes'. Moreover, during the inference stage, constrained text generation is used as we have prior knowledge that the target output is limited to either `yes' or `no'. To ensure the resulting word probability, the predicted probabilities of tokens `yes' and `no' would sum up to 1, and the next token must be `<eos>' as this matches our predefined output.

\section{Experiments}\label{section:experiment}
We conduct experiments using the Microsoft News Dataset (MIND) to evaluate the effectiveness of \model/ and compare it with baseline methods. Our primary goal is to investigate several research questions about the performance of \model/:
\begin{itemize}
\item \textbf{RQ1}: How does \model/ perform compared to other baselines in the task of sequential news recommendation?
\item \textbf{RQ2}: Does \model/ possess the adaptability to construct personalized prompts that incorporate certain knowledge? 
\item \textbf{RQ3}: Is it possible to manipulate \model/ to produce customized recommendations according to the specific needs of users? For instance, suppose a user has certain requirements after reading some articles, such as only wanting to read articles on a similar or different topic from the most recent articles. In such cases, can \model/ adjust itself to meet the users' particular demands and provide relevant recommendations?
\item \textbf{RQ4}: What is the impact of the ranking loss $L_{BPR}$ on the performance of \model/?
\item \textbf{RQ5}: How does the definition of `particular knowledge' influence the performance of \model/?
\end{itemize}

\subsection{Experimental Setup}\label{subsection:setup}
\textbf{Dataset.}
We conduct experiments over the MIND dataset \cite{wu2020mind}, which is a well-established benchmark for researchers in the field of news RS. We utilize data on users' clicked behaviors from November 9, 2019 to November 15, 2019 to examine the effectiveness of treating the news recommendation as a text-to-text language task. The impressions collected from November 9, 2019 to November 14, 2019 are used for model training, while the impressions on November 15, 2019 are used for validation and testing. MIND provides a rich textual information for each news article, including its topic/category, subcategory, title, abstract, title entities, and abstract entities. The specific textual information utilized for news representation depends on the nature of the recommendation task being addressed. A summary of the statistical details of the MIND dataset used in our experiments is provided in Table \ref{tab:MIND}.  

\begin{table*}[t!]
\begin{center}
\caption{Statistics of MIND used for model evaluations.}
\begin{tabular}{c c c c c c c} 
\hline
 \#users & \#news & \#impressions & avg. history length& avg. click rate (\%) &avg. title length& \#category \\ 
 \hline
141,935 & 71,671 & 297,715  & 23.56  & 0.10 & 10.77 & 18\\
 \hline
\end{tabular}
\label{tab:MIND}
\end{center}
\end{table*}

\textbf{Baseline Methods.}
We compare the performance of \model/ with several baseline models using a range of metrics. The baseline models are divided into two groups. The first group includes popularity-based models, which are MostPop and RecentPop \cite{ji2020re}. MostPop measures news popularity based on the number of real-time news clicks, while RecentPop measures it based on the number of real-time news clicks within a recent time. The second group includes personalized deep neural news RS methods, namely LSTUR \cite{an2019neural}, TANR \cite{wu2019neural4} NRMS \cite{wu2019neural2}, and NAML \cite{wu2022news}. LSTUR models a user's long- and short-term interests, TANR trains a topic-aware news encoder via jointly training with a topic classification task; it uses attention network to selection import words from news title and select important news from user's past behavior. NRMS learns users' and articles' representations via multi-head self-attention network, and NAML models users and news articles via multi-view self-attention network. 

\textbf{Implementation Details.} The \model/ proposed in this study utilizes the T5-small pre-trained checkpoint \cite{raffel2020exploring} as its backbone. It comprises 6 layers in both the encoder and decoder components, with a dimension size of 512 and an 8-headed attention mechanism. The SentencePiece \cite{sennrich2015neural} tokenizer is employed, with a vocabulary size of 32,128 sub-word units. Constrained text generation is utilized to generate target templates from the auto-regressive decoder. The batch size is 16. However, in order to incorporate ranking loss $L_{BPR}$ for each sample, a pair of positive and negative sample is generated every time even though the batch size is set to 16. This results in the generation of 32 input-target templates for each batch. To ensure fairness in comparison, for training involving negative sampling, a positive-to-negative sample ratio of approximately 1:4 is employed for the \model/ and all baseline models, and also all baselines are configured to their optimal parameters. For the baseline RecentPop, we use clicked and unclicked impressions in the past 24 hours to calculate the near real-time recent popularity of news articles.

\textbf{Evaluation Metrics.} Various metrics are employed to evaluate models for the task of sequential news recommendation. These metrics include top-$k$ Hit Ratio (HR@$k$), Mean Reciprocal Rank (MRR), and Normalized Discounted Cumulative Gain (NDCG@$k$). Additionally, Gini@$k$ (the Gini index), Topic@$k$, and New@$k$ are also used to measure topic diversity. Gini@$k$ evaluates the overall topic diversity within a list of $k$ recommended articles, while Topic@$k$ calculates the number of distinct topics in the recommended $k$ articles. New@$k$ measures the number of articles in the top $k$ recommendations that are different from previously clicked topics. Enhancing the performance of sequential recommendation typically results in a reduction in the diversity of recommendations. In our study, we aim to enhance the sequential recommendation task using our proposed model \model/, while also ensuring that diversity is not excessively compromised. Therefore, we introduce an additional trade-off score $$\mathrm{tradeoff} = \frac{2\times \nDCG@k\times \Gini@k}{(\nDCG@k + \Gini@k)}$$ 
as the harmonic mean between performance and diversity. For all tables in the following, \textbf{bold} numbers indicate the best performance.

\subsection{Performance Evaluations} \label{section:sequential}
Within this subsection, we first assess the effectiveness of \model/ in the context of sequential news recommendation. Additionally, we explore the capacity of \model/ to integrate particular knowledge into personalized prompts. Lastly, we examine \model/'s ability to provide recommendations that meet particular user requirements, which is also the main advantage of \model/ that distinguish it from the existing baselines. The performances of \model/ and relevant baselines are presented in Table \ref{tab:sequential_cold}, Table \ref{tab:diversity}, and Figure \ref{Fig:control_diverse}.

\subsubsection{Sequential News Recommendations (RQ1)} \label{section:sequential_performance} We assess the effectiveness of \model/ in sequential news recommendations and ensure a fair comparison with other baseline methods by incorporating information on the subcategory and title of the news articles for \model/ and the other methods, i.e., input template (1) and article description (1) from Figure~\ref{fig:template}. For users with a history length shorter than the setting of the history length, \model/ simply adds padding tokens at the end of the input template. On the other hand, existing deep neural news RS add vector embeddings to fill in the remaining articles to a fixed length. As a result, \model/ is capable of considering different lengths of history throughout the training, while the baseline models cannot, which is an advantage of \model/ over existing baselines.

The results presented in Table~\ref{tab:sequential_cold} provide several insights about personalized news RS. Firstly, the methods based on popularity (MostPop and RecentPop) are shown to be more effective than the neural network model LSTUR. This could be attributed to LSTUR's limitations in accurately capturing the user's preferences or the distinct nature of the dataset, where users tend to favor popular articles. Furthermore, the measurement of news popularity could also be influenced by impression bias. Secondly, the performance of our approach, \model/ (1-1), is comparable to other deep neural baselines. This is because our approach also considers users' interests from their historical behaviors, attempts to understand the articles read by a user before, and employs the attention mechanism. Lastly, adjusting the hyper-parameters leads to superior performance of our approach, \model/ (1-1)$^*$, demonstrating the effectiveness of our methodology in introducing ranking loss and utilizing paired data to improve the language model's performance on news recommendation task. Overall, the performance comparison on sequential news RS indicates that treating personalized news recommendation systems as a language task and utilizing constrained text generation with the assistance of  prompt learning is an effective approach in comprehending texts for sequential news RS.

\begin{table*}[t!]
\centering
\caption{The performances of different methods on both personalized news RS and pure cold-start user problems. * indicates that the hyper-parameter $\lambda$ in the training objective is adjusted; otherwise, $\lambda = 0.$ The statistical significance was assessed using the Student's t-test, with a significance level of $p < 0.05$.}
\begin{adjustbox}{width=0.97\linewidth}
\begin{tabular}{c c c c c | c c c c }  
\hline 
\multicolumn{5}{c|}{Performances on sequential news RS} & \multicolumn{4}{c}{Performances on pure cold-start users}\\
 \hline
 Methods &  MRR & HR@5  & NDCG@5 & NDCG@10 &  MRR & HR@5  & NDCG@5 & NDCG@10  \\ 
\hline
MostPop  &0.2699 &  0.4899 &0.2906& 0.3510 & 0.2699 &  0.4899 &0.2906& 0.3510 \\
RecentPop &0.2704  &0.4939 &0.2924  &0.3519 &\textbf{0.2704}  &\textbf{0.4939} &\textbf{0.2924}  &\textbf{0.3519}\\
LSTUR & 0.2522 & 0.4715 & 0.2712 & 0.3352 & 0.2143 & 0.4027 & 0.2258 & 0.2842 \\
TANR &  0.2918 &  0.5519 &  0.3241 & 0.3876 & 0.2180 & 0.3905 & 0.2232 & 0.2855 \\ 
NRMS  & 0.2847 & 0.5253 & 0.3101 & 0.3763& 0.2501 & 0.4600 & 0.2669 & 0.3289 \\
NAML  & 0.2943 & 0.5426 & 0.3235 & 0.3870& 0.2180 & 0.3905 & 0.2232 & 0.2855 \\

\hline
\model/ (1-1) & 0.2924&0.5450&0.3218& 0.3862 & 0.2308 & 0.4458 & 0.2472 & 0.3107 \\
\model/ (1-1)$^*$&  \textbf{0.3084}&\textbf{0.5574}&\textbf{0.3387}&\textbf{0.4012}&0.2469&0.4888&0.2721 & 0.3322 \\
\model/ (2-1) & 0.3048&0.5521& 0.3341& 0.3971 & 0.2602&0.4839&0.2790& 0.3464\\
\hline
\end{tabular}
\end{adjustbox}
\label{tab:sequential_cold}
\end{table*}

\subsubsection{Knowledge-aware Personalized Prompts (RQ2)} Recent studies \cite{cinelli2021echo, lunardi2020metric} highlighted the importance of improving diversity in news RS to enhance the user experience in the long term and mitigate the societal issues arising from echo chambers \cite{cinelli2021echo} and filter bubbles \cite{lunardi2020metric}. Traditional systems that rely solely on users' interests and behaviors tend to provide recommendations that are too similar, potentially harming users' overall experience. In light of this issue, we evaluate the effectiveness of our proposed \model/ and compares it to the baselines that rely on popularity and the top-2 results from the other baselines in Table \ref{tab:sequential_cold}, with a focus on topic diversity. In the following paragraphs, we also assess \model/'s adaptability and flexibility in creating personalized prompts that integrate specific knowledge --- articles' diversity, articles' popularity, and static attributes of users. Our study considers not only \model/'s performance in providing sequential recommendations but also the diversity of recommended topics and the trade-off score mentioned in section~\ref{subsection:setup}.

\textbf{Inclusion of articles' popularity.} 
Table \ref{tab:diversity} presents the results of the baseline methods on recommendation performance and diversity of recommended articles. The results indicate a negative correlation between accuracy, as measured by nDCG@$k$, and diversity, as measured by Gini@$k$. It is evident from the table that methods based on popularity achieve a higher Gini index, indicating that they can recommend a diverse range of top-$k$ articles from various topics. Additionally, these methods also demonstrate higher Topic@$k$ and New@$k$ values, suggesting that popular articles may cover different topics, and as they do not rely on users' previous behaviors to make recommendations, the suggested articles may vary from the user's typical reading preferences. However, as these methods do not consider users' interests, the trade-off between accuracy and topic diversity is lower than that of personalized news RS. This observation motivates us to incorporate the popularity signal of articles into a personalized prompt for \model/ to achieve a better balance between the recommendation accuracy and diversity. 

To account for limitations in the available MIND data, we developed a method for measuring the popularity of articles using the real-time click count. We consider an article to be popular if its real-time click count is above the s$^\text{th}$  percentile of click count for all viewed articles. For articles that do not meet this popularity threshold, we assign the label `personal' to them. We use this information in conjunction with the input template (1) and article description (2) shown in Figure \ref{fig:template} to train our model. Our study demonstrates that our \model/ (1-2)$^*$ performs better than popularity-based and deep neural news RS baselines in terms of nDCG@$k$ and the trade-off score.

\textbf{Inclusion of articles' diversity.} Following a comprehensive examination of the dataset, it was discovered that 54\% of the 297,715 impressions contained articles on new topics that were clicked, which were distinct from the topics of articles used to construct users' historical models. This discovery motivated us to enhance the article descriptions by introducing a signal of diversity. An article is classified as diverse if its topic differs from that of the $T$ most recent articles. Otherwise, it would be categorized as `personal'. To incorporate an indicator that signifies the diverse nature of the article, we utilize the input template (1) and the article description (3) from Figure \ref{fig:template} in our model training. Similar to the findings obtained by incorporating popularity signals in article descriptions, the inclusion of diversity signals in articles can improve the performance of the \model/ (1-3)$^*$ model compared to the current deep neural news RS and \model/ (1-1)$^*$ that does not utilize the diversity signal.

\begin{table*}[t!]
\centering
\caption{Comparisons on recommendation performance and topic diversity in recommended list. * indicates that the hyper-parameter $\lambda$ in the training objective is adjusted; otherwise, $\lambda = 0.$ The statistical significance was assessed using the Student's t-test, with a significance level of $p < 0.05$.}
\resizebox{\textwidth}{!}{\begin{tabular}{c c c c c c c c c c c } 
 \hline
 Methods &  HR@5  & NDCG@5 & NDCG@10 & Gini@5 & Gini@10 & Topic@5 & Topic@10 & New@5 & New@10 & tradeoff\\ 
 \hline
 MostPop& 0.4899&0.2906& 0.3510&\textbf{0.6716}&\textbf{0.7675}&\textbf{3.6178}&5.8699&\textbf{2.7593}& \textbf{5.4906}&0.4437\\
 RecentPop& 0.4939& 0.2925& 0.3519 & 0.6695& 0.7664 & 3.5924 &\textbf{5.8734} & 2.7331 &5.4630 & 0.4447\\

TANR & 0.5519 &  0.3241 & 0.3876 & 0.5934 & 0.7073 & 3.2685 & 5.1822 & 2.3216 & 4.6841 &0.4601\\
NAML & 0.5426&0.3235&0.3869 & 0.5989& 0.7109 & 3.2958 & 5.2412 & 2.2952 & 4.6713 & 0.4605\\
\hline
\model/ (1-1)$^*$ & 0.5574 & 0.3387 & 0.4012 & 0.5678 & 0.6972 & 3.1467 & 5.1722 & 1.7172 & 3.9584 &\textbf{0.4668} \\
\model/ (1-2)$^*$ & \textbf{0.5714} & \textbf{0.3470} & \textbf{0.4076} & 0.5384 & 0.6794 & 2.9884 & 4.9450 & 1.6103 & 3.7911 & 0.4658 \\ 
\model/ (1-3)$^*$ & 0.5688& 0.3454& 0.4068 &0.5464&0.6828&3.0369& 4.9535&1.6092& 3.7424 &0.4666 \\

 \hline
\end{tabular}}
\label{tab:diversity}
\end{table*}

\textbf{Inclusion of users' static attributes.} The cold start problem happens when a user has no historical behavior (i.e., cold-start user) \cite{lam2008addressing} or an item is new (cold-start item)  \cite{schein2002methods}. In news RS, the cold-item problem can be alleviated by representing each article using its textual information. However, it is difficult for sequential news RS to make recommendations when there is no available user past behaviors. We evaluate the performance of models trained for personalized news RS in section \ref{section:sequential_performance} in recommending news to users who have not clicked on any articles before (i.e., pure cold-start users). The experimental results in Table \ref{tab:sequential_cold} yield several noteworthy findings. First, we find that popularity-based models consistently outperform all personalized methods. This can be attributed to the fact that personalized methods rely on users' interests to make recommendations, which poses a significant challenge when attempting to model the interests of cold-start users. In contrast, popularity-based methods recommend news based on news popularity, which often correlates with article importance, such as in the case of an earthquake. As a result, readers are more likely to click on and read articles that they find important or relevant, regardless of their personal interests. The second finding indicates that the trained \model/ outperforms the trained deep neural news recommendation system baselines in providing recommendations to users without any historical behavior. This could be attributed to the fact that \model/ takes into account varying lengths of historical user behaviors during training and users with sparse past reading behavior are included in the training dataset.

When trying to provide useful news recommendations to users who have no browsing history, it can be advantageous to use demographic information such as age, gender, and location from other domains \cite{lam2008addressing}. However, incorporating this information into personalized neural network news recommendation methods requires the development of a complex model that can analyze the user's profile, which may necessitate changes to the architecture of the system. Alternatively, our proposed approach, called \model/, simplifies the process by modifying personalized prompts. This model focuses only on a user's short-term interests, which can change frequently. However, many online news platforms categorize their articles into distinct topics, which can capture a user's long-term interests. By incorporating this attribute into our model, we can potentially improve the effectiveness of personalized news recommendations. We trained a model called \model/ (2-1) using prompts that consider both a user's recent behavior and their interests in specific topics. Our results show that \model/ (2-1) outperforms other personalized news recommendation systems that do not use profile information, although it still falls behind popularity-based models.

The preceding experiments demonstrate that by integrating knowledge into personalized prompts, \model/ can enhance its flexibility to effectively incorporate additional textual information into article descriptions or model users' interests without altering the underlying model architecture and the training objective.

\subsubsection{Controllability of \model/ (RQ3)}
The present news recommendation models are designed for specific purposes, and any modifications to the article descriptions could require changes to the model as well. In this study, we demonstrated the flexibility and adaptability of our model by incorporating additional details into article descriptions and user interests. Our objective now is to investigate the level of control we have over our model, which is its main strength, in comparison to the existing news RS.

We define a RS as controllable if it can generate recommendations based on individual users' needs. It is essential to have control over the news recommendations we receive because after reading some articles, users may be interested in exploring topics that differ from those they have previously read. Since users have different reading habits, there must be a way for them to communicate their preferences to the RS, enabling it to generate recommendations that are personalized to their specific interests. For instance, a user who has read numerous articles on one topic may prefer to read an article on another topic next. Current news RS may not identify this preference and continue recommending articles from the same topic. We now develop a model that can consider users' requests for diverse topic articles and enhance the news RS accordingly.

Our study aims to explore whether \model/ can generate personalized recommendations based on users' specific requirements. To achieve this, we employ input (3) as our input template and use description (3) as shown in Figure \ref{fig:template} to assess the controllability of \model/ in recommending articles that are tagged as `diverse'. The training data must be adjusted for the study of \model/'s controllability, where positive samples are consistently identified as `diverse', but negative samples may labelled as either `diverse' or `personal'. Similarly, we also use input template (4) and article description (3) to evaluate whether \model/ can provide personalized recommendations when necessary, with a positive-to-negative sample ratio of approximately 1:4. To demonstrate \model/'s effectiveness in considering users' preferences in generating recommendations, we test its performance on three groups of testing impressions: (1) all impressions in the test dataset, (2) diverse impressions in the test set where all clicked articles are labeled as `diverse', and (3) personal impressions in the test set where all clicked articles are labeled as `personal'.

\begin{figure}
     \centering
     \begin{subfigure}[b]{0.45\textwidth}
       \centering
		\includegraphics[width = 1\textwidth]{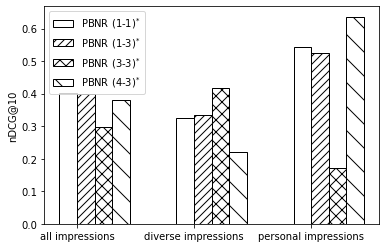}
		\caption{Model performances on sequential news recommendation task.}
     \end{subfigure}
     \hfill
     \begin{subfigure}[b]{0.45\textwidth}
         \centering
        \includegraphics[width=1\textwidth]{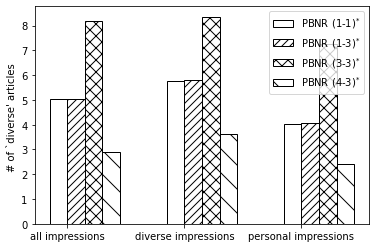}
         \caption{The number of articles labelled as `diverse' within the top-10 recommendations.}
     \end{subfigure}
     \caption{Evaluation of \model/'s controllability to make recommendations based on individual user requirements, which can
enhance user experiences, improve human-computer interaction, and to some extent improve the interpretability of news RS. \model/ (1-1)$^*$ and \model/ (1-3)$^*$ do not consider users' requirements, while \model/ (3-3)$^*$ aims to recommend articles labelled as `diverse' and \model/ (4-3)$^*$ aims to recommend articles labelled as `personal'.} 
     \label{Fig:control_diverse}
\end{figure}

Figure \ref{Fig:control_diverse} compares the performances among different models for news recommendation. Subfigure (a) demonstrates that \model/ (1-1)$^*$ and \model/ (1-3)$^*$ perform similarly in providing sequential news recommendations, while \model/ (3-3)$^*$ and \model/ (4-3)$^*$, which aim to recommend articles based on users' preferences for topic diversity, perform worse than \model/ (1-1)$^*$ and \model/ (1-3)$^*$. However, \model/ (3-3)$^*$, which targets articles labeled as `diverse', performs better than all other models in terms of diverse impressions, while \model/ (4-3)$^*$, which targets articles labeled as `personal', performs better than all other models in terms of personal impressions. Subfigure (b) also presents the number of recommended articles labeled as `diverse' among the top-10 recommendations. As expected, \model/ (3-3)$^*$ suggests a more varied range of topics, while \model/ (4-3)$^*$ recommends a limited range. These findings confirm the adjustability of \model/, which enables one to tailor it to provide either personal or diverse recommendations based on readers' preferences. The findings confirm the controllability of \model/, as well as to some extent improve the interpretability of news RS. This is beyond the capability of currently existing RS that are designed for news recommendation.

\subsection{Ablation Study on Ranking Loss (RQ4)} \label{section:ablition_loss}

\begin{figure}[h]
\centering
\includegraphics[width=7cm]{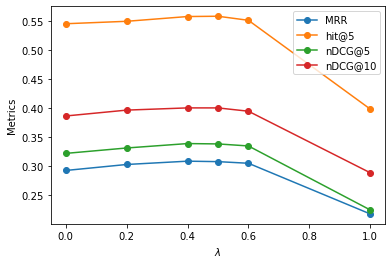}
\caption{\model/ performance on sequential recommendation with different $\lambda$ values -- weight on ranking loss.}
\label{fig:RQ2}
\end{figure}

This section describes an ablation study of the training objective function to assess the impact of jointly training the ranking loss $L_{BPR}$ and the language generation loss $L_{NLL}$ in the training process. We expect the model to distinguish between a user's preference for clicked and non-clicked articles, while also providing relevant recommendations via language generation. The results are presented in Figure \ref{fig:RQ2}. We find that not considering $L_{BPR}$ results in suboptimal recommendations. If $\lambda$ is too small (or equals 0), the model fails to fully utilize the benefits of adopting $L_{BPR}$. Conversely, if $\lambda$ is too large (or if we solely focus on $L_{BPR}$), the performance of the language model in generating responses may be overlooked, leading to a decline in overall performance. This observation highlights the significance of taking the ranking loss into account during training to enhance the language model's performance on recommendation task.

\subsection{Influence of Definitions of `Particular Knowledge' (RQ5)}
Section \ref{section:sequential} showed that the \model/ can effectively integrate additional knowledge to enhance its performance. This section presents our experimental analysis of the impact of specific knowledge definitions on the model's recommendations.

\begin{figure}
\centering
\begin{subfigure}[b]{0.45\textwidth}
\centering
\includegraphics[width=6cm]{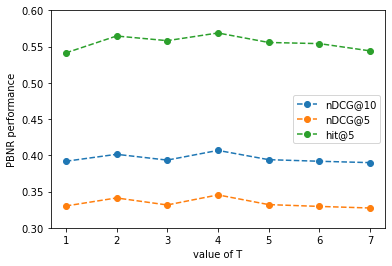}
\caption{Recommendation performance with the knowledge of articles' diversity.}
\end{subfigure}
\hfill
\begin{subfigure}[b]{0.45\textwidth}
\centering
\includegraphics[width=6cm]{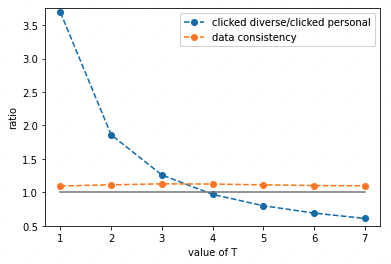}
\caption{Visualization of training data samples and the consistency between the training data and the testing data. }
\end{subfigure}
\caption{Evaluation of \model/ performance on recommendation with different definitions of diversity.}
\label{Fig:influence_diversity}
\end{figure}

\begin{figure}
\centering
\begin{subfigure}[b]{0.45\textwidth}
\centering
\includegraphics[width=6cm]{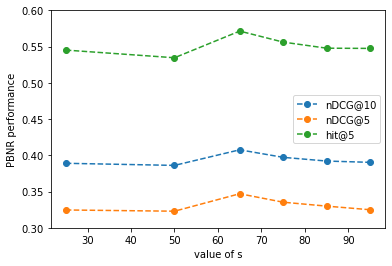}
\caption{Recommendation performance with the knowledge of articles' popularity.}
\end{subfigure}
\hfill
\begin{subfigure}[b]{0.45\textwidth}
\centering
\includegraphics[width=6cm]{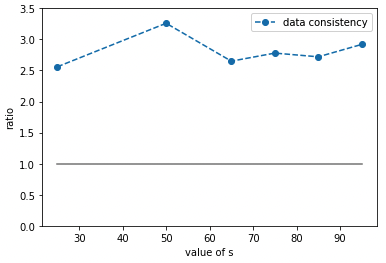}
\caption{The consistency between the training and the testing data.}
\end{subfigure}
\caption{Evaluation of \model/ performance on recommendation with different definitions of popularity.}
\label{Fig:influence_popular}
\end{figure}

We define an article as `diverse' if its topic differs from those of the $T$ most recently read articles in the user's history; otherwise, it is considered `personal'. Figure \ref{Fig:influence_diversity} illustrates the performance of \model/'s recommendations with various values of $T$. Subfigure (a) shows a general trend that the performance decreases as $T$ either increases or decreases, and we observe that $T = 4$ is an appropriate choice for defining articles' diversity to achieve the best recommendation performance. To incorporate the popularity of articles into the model, we define an article as `popular' if its click count in real-time exceeds the s$^\text{th}$ percentile of all viewed articles. We observe that the recommendation for such articles declines as the value of $s$ increases or decreases, and that the optimal performance is achieved when $s = 65$, as depicted in Figure \ref{Fig:influence_popular}.

One possible explanation for the findings is the memorization ability of the large language model. To assess the influence of article diversity, we analyzed the proportion of articles labeled as `diverse' versus `personal' that were clicked on, denoted as clicked diverse/clicked personal. Based on subfigure (b) in Figure \ref{Fig:influence_diversity}, our results demonstrate that when $T$ equals 4, the proportion of clicked articles labeled as `diverse' is approximately equal to those labeled as `personal', indicating no dominant label during the training process. This observation implies that the language model may memorize the `diverse' signal when generating the output sequence for the testing data. Optimal performance of the \model/ is achieved when the memorization capability of the large language model is reduced. 

Similarly, when considering the popularity label of articles as an additional prompt, we observe that either the `personal' or `popular' label consistently dominates during the training phase, indicating the activation of the language model's memorization capacity. To ensure that the `popular' signal is properly utilized during keyword memorization, it is crucial to assess whether the testing data follows a similar pattern as the training data. We evaluate this consistency by calculating the ratio between the number of clicked articles labeled `popular' and `personal' in both training and testing datasets. A ratio between these two ratios closer to 1 indicates a higher level of consistency between the two datasets in subfigure (b) of both Figure \ref{Fig:influence_diversity} and Figure \ref{Fig:influence_popular}. Our results, illustrated in subfigure (b) of Figure \ref{Fig:influence_popular}, indicate that the model performs optimally when the testing data follows a similar pattern as the training data while the memorization capacity is activated.

Since language model show its capability of memorization, when incorporating additional textual information to enhance the model's performance, it is crucial to carefully define this particular knowledge.

\section{Conclusion}\label{section:conclusion}
In this work, we introduce a novel news recommendation approach called \model/ that capitalizes on the strengths of pre-trained language models and prompt learning. Rather than considering news recommendation as a conventional task, we treat it as a text-to-text language task. To improve the language model's performance for the recommendation task, we incorporate both ranking and language generation losses during model training. Our experimental findings show that \model/ outperforms current baselines in recommendation accuracy and does not require a fixed length of history for all users throughout the training process. This improvement can be attributed to the enhanced language understanding capabilities of pre-trained language models. Unlike other baselines that may necessitate a change in the model's architecture to integrate additional information, \model/ remains unchanged in structure and training objectives, and extra information can be easily incorporated through prompt design. \model/ also stands out from other deep news RS methods in its ability to produce customized recommendations to meet users' specific needs, improving the human-computer interaction in the domain of news RS through the memorization capabilities of large language models.

We must recognize and address the limitations of the present study. One limitation involves the use of only news titles and subcategories as a means of representing articles. The MIND dataset has additional information, including article bodies and entities, which are essential for comprehending an article's content. To improve the performance of the system, future research should consider incorporating news entities or additional textual information in prompts within the constraint of input token limits. Moreover, our study employed hard prompts, and manually designing personalized prompts is a time-consuming process. Possible future research could develop automated approaches to prompt design, which would allow the system to design prompts more efficiently and independently.

%\catcode`'=9
%\catcode``=9
\bibliographystyle{ACM-Reference-Format}
\bibliography{ref}

\end{document}